\documentclass{article}
\usepackage{graphicx} % Required for inserting images
\usepackage[square,numbers]{natbib}
\usepackage[table]{xcolor}
\usepackage{wrapfig}
\usepackage{longtable}
\usepackage{sidecap}
\usepackage{amssymb}
\usepackage{geometry}
\newlength{\colwidth}

\usepackage[table]{xcolor}
\definecolor{lightgray}{gray}{0.9}  % For alternating row colors in tables

\title{X-ray grating spectroscopy as a mission enhancement}

\author{Hans Moritz Guenther (MIT)\footnote{hgunther@mit.edu}
\and Ehud Behar (Technion, Israel)
\and Joel N. Bregman (Univ. Michigan)
\and Laura W. Brenneman (Center for Astrophysics | Harvard \& Smithsonian)
\and Alexander R. Bruccoleri (Izentis)
\and L\'{i}a Corrales (University of Michigan)
\and Elisa Costantini (SRON, The Netherlands)
\and Thomas Dauser (FAU Erlangen-Nuernberg)
\and Casey T. DeRoo (U. Iowa)
\and Abraham D. Falcone (Penn State University)
\and Adam R. Foster (Center for Astrophysics | Harvard \& Smithsonian)
\and Luigi Gallo (Saint Mary's University)
\and Catherine E. Grant (MIT)
\and Sean J. Gunderson (MIT)
\and Ralf K.~Heilmann (MIT)
\and David P.~Huenemoerder (MIT)
\and Maurice Leutenegger (NASA/GSFC)
\and Eric D.~Miller (MIT)
\and Michael Nowak (WashU)
\and Frits Paerels (Columbia)
\and David A. Principe (MIT)
\and Ioanna Psaradaki (ESA)
\and Andrew Ptak (NASA/GSFC)
\and Agata Rozanska (NCAC PAS, Poland)
\and Randall K Smith (Center for Astrophysics | Harvard \& Smithsonian)
\and Pasquale Temi (NASA/ARC)
\and Todd M. Tripp (University of Massachusetts)
\and Lynne Valencic (JHU, NASA/GSFC)
\and Joern Wilms (FAU Erlangen-Nuernberg)
\and Scott J. Wolk (Center for Astrophysics | Harvard \& Smithsonian)
}
\date{For CoPAG, PhysPAG}

\begin{document}

\maketitle

%\todo{Structure follows the ASTRA Mission Concept Description Template V2.1. See that template to see what is expected in each section. While it's not said there directly, the general tone of the instructions is``short is better'', so I'm actually quite happy with, e.g., the mission section being shorter than the limit. Page limits are per section. I'll adhere to those limits, but for readbility, the sections might not break exactly that the end of a page.}

% \newpage

\section{Science Investigation} % 1-2 pages
\label{seft:science}
%The life cycle of baryons through the universe is one of te fundamental open questions in astronomy. In the NASA roadmap ``Enduring quests, daring visions'' from 2014 it is called out as the ``stellar life cycles and the evolution of elements'', while the more recent 2020 decadal survey incorporates this question in two of its three priority scientific areas, split by the class of objects observed: ``Pathways to habitable worlds'', which requires understanding of the elements that make up those potentially habitable worlds and ``Drivers of galaxy growth'' that depends on the content and feedback of the interstellar and intergalactic medium. While progress has been made since those reports, in particular through new observations with JWST, the questions of where those baryons are and how they cycle in and out of stars and galaxies is still very much open (Fig.~\ref{fig:Habex_baryon_lifecycle}).
Feedback from accreting black holes regulates galaxy growth,
operating across an enormous range of physical scales.
A supermassive black hole (SMBH) launches winds at $\lesssim 0.01$~pc, drives that energy through the host galaxy on kpc scales, and ultimately shapes the hot
gas in galactic halos and the intergalactic medium.
Both the 2014 NASA roadmap ``Enduring quests, daring visions'' and the 2020 Decadal Survey prioritize these physics. Astro2020's ``Drivers of galaxy growth'' seeks to quantify how feedback shapes gas content across cosmic time, while ``Pathways to habitable worlds'' requires a census of the elements that build planetary systems. Tracing feedback across these scales reaches into three baryon populations that current telescopes cannot observe: the hot winds from SMBHs, the hot halo gas that holds most of the universe's ``missing'' baryons, and, within galaxies, metals bound in dust gains. All three demand deep, high-resolution soft X-ray spectroscopy.

\begin{figure}
%    \centering
    \includegraphics[width=0.5\textwidth]{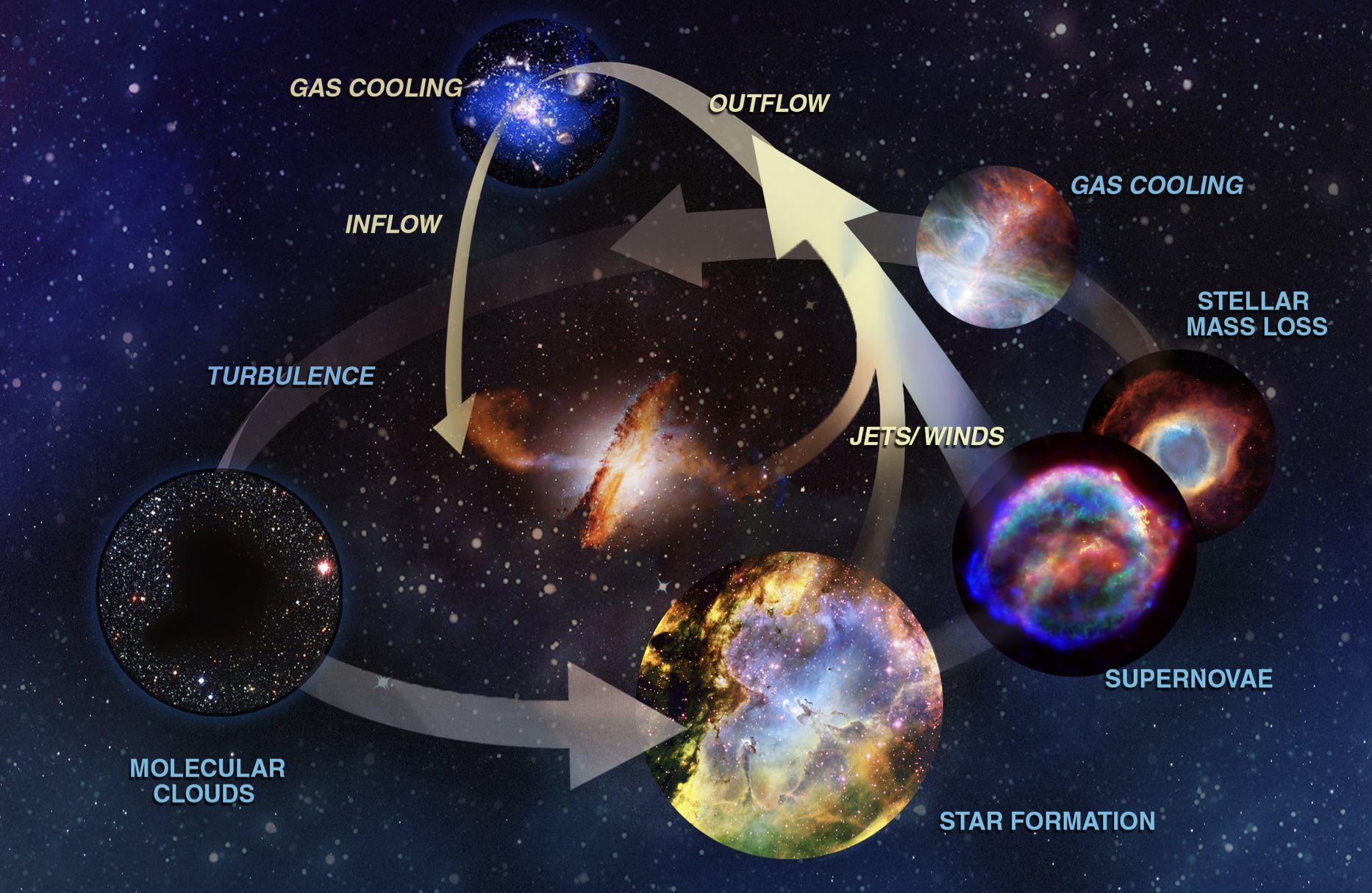}
    \includegraphics[width=0.45\textwidth]{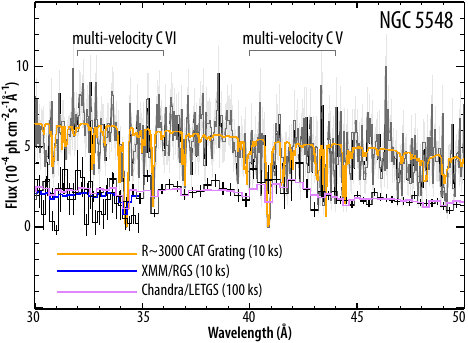}
    \caption{[Left] Following the lifecycle of baryons in galaxies requires high-resolution soft X-ray spectroscopy to complement upcoming missions like HWO and existing missions such as HST and JWST that miss baryons in hot phases. X-ray emission and absorption lines quantify metals in the hot gas of the ISM, in AGN winds, or in stars while X-ray absorption edges can determine abundances and structure of dust grains. Image source: HabEx Report, The Habitable Exoplanet Observatory Study Team. [Right] Simulated spectra of the AGN NGC 5548 with a high-performance grating compared to Chandra and XMM results.}
    \label{fig:Habex_baryon_lifecycle}
\end{figure}

\subsection{AGN Wind Outflows and Feedback}

Over cosmic timescales, SMBHs regulate star formation via accretion-driven winds. The SMBH mass correlates tightly with the velocity dispersion of its host bulge
\citep{1998AJ....115.2285M, 2000ApJ...539L...9F},
a connection that demands energy transfer from the accretion scale ($\lesssim 0.01$~pc) to the galaxy scale (kpc and beyond). X-ray winds carry the bulk of this feedback.
The hot, highly ionized gas seen in X-ray absorption carries up to $10^3$ times the column density of the cooler UV outflows that HST has characterized, and far more
than the optical-line gas \citep{2015ARA&A..53..115K, 2010A&A...521A..57T}. That same outflow reappears downstream as the entrained neutral and molecular gas that ultimately leaves the galaxy or quenches star formation. A complete accounting of feedback thus requires the soft X-ray band, where the dominant ionization states of the launched wind are found.

Quantifying AGN feedback requires the wind's mass outflow rate and kinetic power. %, both of which depend on the gas density at the launch point. Current X-ray telescopes cannot measure this density. 
The deepest Chandra and XMM-Newton grating exposures of the brightest AGN resolve multi-component winds in velocity and ionization, but the density-sensitive diagnostics lie below their threshold for all but the nearest obscured AGN. X-ray microcalorimeters lack the necessary spectral resolution at these soft energies, and XRISM/Resolve cannot reach them at all without an open gate valve.
%Even with one, a calorimeter delivers $R \approx 2000$ only above $\sim2$~keV; below that energy, where O~{\sc vii}, O~{\sc viii}, and the density-diagnostic ions reside, grating spectroscopy provides several times the resolution.
A grating spectrometer with $R > 3000$ and effective area several times that of Chandra resolves the density, distance, and kinematics of these winds across a representative AGN sample (see Fig~\ref{fig:Habex_baryon_lifecycle}[Right]), linking accretion-scale feedback to its galaxy-scale consequences.

%One key problem is that existing observatories miss two phases of baryons, which require deep high spectral resolution X-ray spectroscopy as described in the next two sections.

\subsection{Feedback on Halo Scales: the Circumgalactic Medium and the Missing Baryons}

The hot gas in galactic halos is the leading candidate reservoir for the universe's missing baryons, and its distribution records the integrated history
of feedback. Massive galaxies contain less than one-quarter of the baryons expected from the cosmic baryon fraction \citep{2013A&A...557A..52P, 2011ApJ...737...22A, 2015JATIS...1d5003B, 2004ApJ...616..643F}, 
and the deficit must reside in a hot, diffuse halo at temperatures $T \sim 10^{5.5}$--$10^7$~K. The dominant diagnostic transitions for this temperature range---O~{\sc vii}, O~{\sc viii}, C~{\sc vi}, and Ne~{\sc ix}---all fall in the soft X-ray band. Absorption-line spectroscopy against bright background AGN provides the only practical probe, as its emissivity is negligible at these densities.

%The shape of the halo density profile is itself a direct measurement of feedback. 
Supernova-, stellar wind-, and AGN-driven energy injection flattens the hot gas distribution relative to the dark matter, redistributing baryons and metals to large radii.
High-resolution X-ray absorption from this CGM gas against background AGN traces the gas's radial density and temperature profile beyond the virial radius. This measurement distinguishes between cosmological simulations whose feedback prescriptions disagree on the halo gas mass of $L^*$ galaxies by a
factor of three \citep{2020MNRAS.491.4462D}. Current instruments detect this absorption only toward the brightest sightlines and only in the inner halo \citep{2022AJ....163..264H}. Mapping the full profile requires an order-of-magnitude gain in sensitivity,
exactly the capability a grating spectrometer with $R > 3000$ provides.

The Milky Way offers the best-characterized hot halo, because every extragalactic sightline passes through it. Resolved O~{\sc vii} and O~{\sc viii} absorption-line profiles measure not only the Galactic hot gas mass but also the halo rotation and the net inflow or outflow rate, quantities presently uncertain by factors of two or more \citep{2017ApJ...849..105L}. Galaxy groups and clusters, where O~{\sc viii} and Fe~{\sc xvii} dominate, extend
this science to the hotter end of the baryon temperature function and probe the cosmic web filaments that thread the intergalactic medium.

% RKS Note: As much as I love dust studies, this section could be trimmed if needed.  I would keep the figure regardless, however.

\subsection{Elements bound in dust grains}
X-ray spectroscopy can quantify interstellar dust, using absorption edges and shapes to track both the elemental abundances and overall composition.

\begin{figure}
    \includegraphics[width=0.49\textwidth]{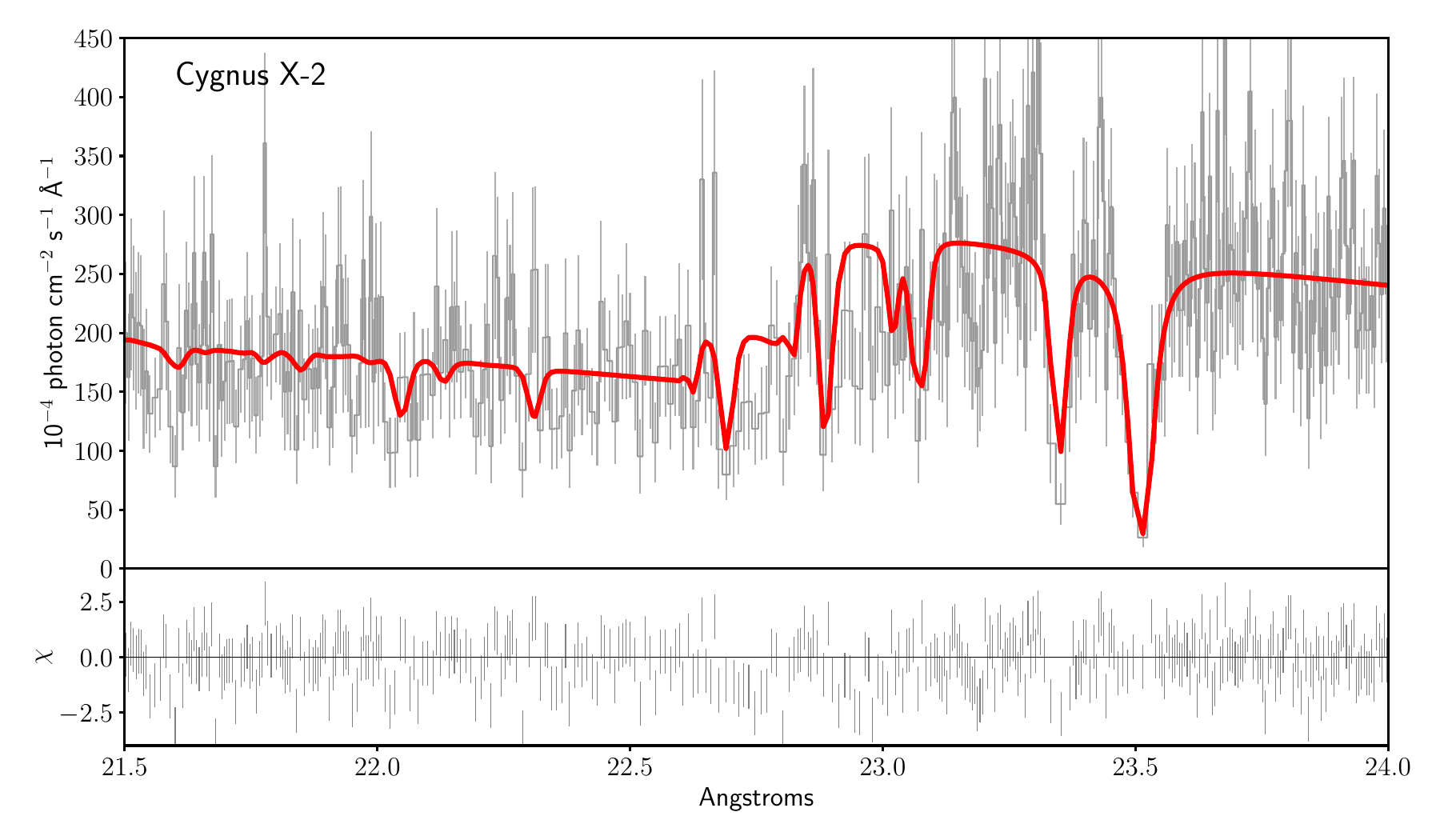}
    \includegraphics[width=0.49\textwidth]{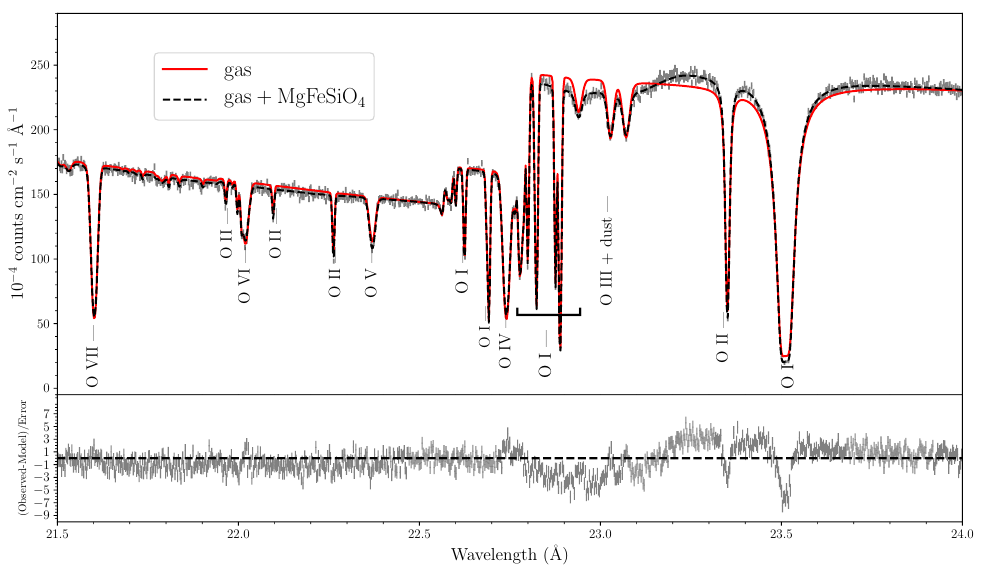}
    \caption{
    [Left] A Chandra MEG spectrum of Cygnus~X-2 (30~ks), showing the O~K-shell absorption complex for interstellar oxygen.
    [Right] A simulated high resolution spectrum of Cygnus~X-2 with an $R \sim 3000$ gratings instrument (modified from Fig.~9 of \cite{2020A&A...642A.208P}).
    % assuming 200 ks with about 400 cm^2 (Arcus response matrixes)
    With high-resolution spectroscopy of this type, residual absorption by O-bearing interstellar dust can be discerned from gas, and its chemical composition can be identified.
    }
    \label{fig:Psaradaki_2020}
\end{figure}

Oxygen is one of the key contributors to the chemistry of the ISM and yet at least one quarter of it cannot be seen (``missing oxygen problem''). NIR spectroscopy helps \cite{2025prim.book..446O} but only X-ray observations of the oxygen K edge can measure all of it. Metals in both gaseous and solid phases imprint absorption edges observable with current telescopes when viewed in front of bright background sources (Fig.~\ref{fig:Psaradaki_2020}). The energy of the edge differs between single atoms and atoms bound in molecules and the structure of the chemical bond leads structure near the absorption edge (XANES),
%In addition, the structure of the chemical bond leads to X-ray absorption near edge structure (XANES),
%also known as near edge X-ray absorption fine structure (NEXAFS), 
%which is routinely used in synchrotron facilities on Earth to determine concentrations and bonds of trace elements. 
but the spectral resolution and S/N of current X-ray telescopes is too limited to use this diagnostic \cite{2018A&A...609A..22R} except for a handful of the brightest sources \cite{2026ApJ..1002...86R}. Those limited studies reveal unique insight into the dust, e.g., they identify Mg-rich amorphous pyroxene and measure the fraction of Fe in a metallic state \cite{2023A&A...670A..30P}. The proposed grating will make such measurements routine for many sightlines throughout our galaxy and enable the study of distributions of dust species with distance from the galactic center. This insight into the chemical evolution of our galaxy is not available by any other means.

%\begin{SCfigure}
%    \includegraphics[width=0.44\textwidth]{ASTRA_OK_figure.pdf}
%    \caption{
%    (Top) A Chandra MEG spectrum of Cygnus~X-2 (ObsID 1102), showing the O~K-shell absorption complex for interstellar oxygen. The spectrum has been binned to a minimum of 9~counts per bin. This observation was taken in 1999, early in the life of the mission. Studying the O~K-shell region with HETG is no longer feasible due to a build up of contaminating material that has reduced the effective area of the HETG-MEG to effectively zero in this bandpass. XMM-Newton RGS still provides a window on O~K, but at lower resolution (40~m\AA) compared to Chandra MEG (23~m\AA). 
%    (Bottom) A simulated high resolution spectrum of Cygnus~X-2 with an $R \sim 3000$ gratings instrument (modified from Fig.~9 of \cite{2020A&A...642A.208P}. With high-resolution spectroscopy of this type, residual absorption by O-bearing interstellar dust can be discerned from gas, and its chemical composition can be identified.
%    }
%    \label{fig:Psaradaki_2020}
%\end{SCfigure}

\subsection{A glaring gap in spectroscopic capabilities in X-rays}
Modern astrophysics relies upon multiband astronomical data with high spectral resolution, available in the radio (ALMA: $R>10^6$), far IR (Herschel/HIFI or SOFIA/upGREAT $R>10^7$), the near-IR and optical (VLT, Keck instruments reach $R\approx 10^5$), to UV (HST/STIS: $R\approx 10^5$). However, in X-rays the best resolution available below 4~keV is $R\approx 1000$ with the aging low-area Chandra/HETG. Calorimeter resolution (e.g., XRISM Resolve, NewAthena X-IFU) \emph{falls} toward lower energies as their $\Delta E \sim$\,constant, so $R = E/\Delta E$ degrades as $E$ drops. For the soft-band science that defines feedback studies, gratings are not a convenience but a necessity. Resolving the thermal broadening of O~{\sc vii} emission from $10^6$~K gas requires $R \approx 10^4$, and measuring absorption-line centroids at $\lesssim 30$~km~s$^{-1}$, essential for the kinematics of both AGN winds, halo gas, and stellar activity, requires $R > 3000$.

%However, in X-rays the best resolution available at energies below 4 keV is $R\approx 1000$ with the Chandra/HETG. Microcalorimeters like XRSIM/Resolve or Athena/X-IFU reach up to $R=2000$ with harder X-rays, while the concept proposed here can reach $R>5000$ in the 0.2-2keV range. Resolving the thermal width of emission lines requires $R\approx 10^4$ for gas hot enough to emit X-rays -- an order of magnitude above current missions. Absoprtion features can be even narrower.

% \newpage

\section{Science Table} % <1 page
This instrument is meant to be part of an observatory for the entire astronomical community, and can address a wide range of science questions beyond the discussion in the previous section. Some examples are listed in the lower part of the table.
\setlength{\colwidth}{3.7cm}
\begin{table}[h!]

\centering
\rowcolors{3}{lightgray}{white}
\begin{tabular}{|p{\colwidth}|p{\colwidth}|p{\colwidth}|p{\colwidth}|}
\hline
\textbf{Science Objectives} & \textbf{Physical Parameters} & \textbf{Observables} & \textbf{Potential Challenges} \\

\hline

\multicolumn{4}{|c|}{\textbf{Science investigation (section~\ref{seft:science})}}\\
\hline

AGN wind feedback: how SMBHs regulate their host galaxies
& Mass-outflow rate, kinetic power, velocity, and ionization vs.\ Eddington fraction
& Absorption-line centroids, widths, and columns; density-sensitive line ratios; variability
& $R > 3000$; area a few$\times$ Chandra \\
%Determine black hole feedback on surroundings & mass, energy, momentum of winds from inner regions around black holes & width and centroid of absorption lines, time evolution & $R > 2000$, effective area a few times Chandra\\

Missing baryons in the hot halos of galaxies, groups, and clusters
& Column density, temperature, metallicity, and gas-mass fraction vs.\ impact parameter
& K$\alpha$ absorption (O, C, Ne, Fe~{\sc xvii}): depth, width, centroid; radial profiles to $>R_{200}$
& $R > 3000$; area a few$\times$ Chandra \\
%Measure the spatial and temperature distribution of hot gas at and beyond the virial radii of galaxies and clusters & column density and temperature as a function of radial distance from nearest galaxy, group, or cluster center & absortion line depth, width, and centroid for K$\alpha$ lines of C, N, O, Ne and some L-shell lines & needs $R > 3000$ and effective area a few times Chandra\\

Structure and dynamics of the Milky Way hot halo
& Density profile to $R_{200}$; rotation; net inflow/outflow rate
& O~{\sc vii} and O~{\sc viii} absorption-line profiles along many sightlines
& $R > 3000$; $\lambda$-calibration $<0.4$~m\AA \\
%\rowcolor{green!10}
%Constrain galactic feedback through determination of mass loss rates in OB Stars & Wind velocity and absorption & Line profiles, moments, fluxes & Sample suitable for current instruments is too small. \\

The missing oxygen problem and ISM dust composition
& Gas/dust oxygen partitioning; grain species (pyroxene, oxides, ices) vs.\ Galactic radius
& O~K edge shape and fine structure (XAFS); C, N, O, Ne, Fe edge depths
& $R > 3000$ to resolve edge structure \\
%Solve ``missing oxygen problem'' in the ISM & (relative) abundances for C, O, Fe in gas vs dust & Distinguish shape of O absorption edge between solids and molecular & Separate different ISM phases with different kinematics. Background sources used as light bulbs may have intrinsic features. Needs $R > 3000$\\

\hline
\rowcolor{white}
\multicolumn{4}{|c|}{\textbf{A few examples for general observatory science}}\\
\hline
\rowcolor{blue!10}
Accretion and the formation of stars and planetary systems
& Accretion rate; shock velocity; abundances in streams and disks
& Line fluxes and profiles; He-like triplet ratios; O~K edge toward disks
& $R > 3000$ \\
\hline 
%How do accretion processes affect the growth and formation of stellar and planetary systems? & mass accretion rates and elemental abundances in accretion streams & line fluxes, line profiles & need $R > 3000$ \\

\rowcolor{green!10}
Exoplanet atmospheric structure
& Density and composition of the upper atmosphere
& Transit depth vs.\ wavelength and elemental edge
& Area to map several edges per transit \\
%\hline
%\rowcolor{blue!10}
%Determine structure of exoplanet atmospheres & density profile of atmosphere & transit depth as a function of wavelength & current effective area allows this for one planet in one band only\\

\rowcolor{blue!10}
Stellar winds and feedback from OB stars
& Wind velocity and mass-loss rate
& \raggedright Wind-line profiles, moments, and fluxes (UV/X-ray)
& $R > 3000$ \\

%Determine oxygen content of proto-planetary disks & abundance of oxygen in grains  & depth of oxygen absoprtion edge & need more collecting area for sufficient S/N \\

%\rowcolor{blue!10}
%understand nucelar burning on WD as SN Ia progenitors  & motion in atmosphere, optical depth & line shift and line shape in absorption lines & line blends are common\\

%\rowcolor{green!10}
%another example & another parameter & another observable & another challenge - do we have anyting that requires $R >> 3000$?\\

\hline
\end{tabular}
\end{table}

\newpage

\section{Instrument Description} % < 1 page
We envision a suite of next-generation spectrographs that can be added to almost any X-ray mission with focussing optics; the primary mission could be geared towards X-ray imaging, timing, polarimetry, or spectroscopy. In either case, we propose to add an X-ray grating spectrometer to vastly extend the host mission's capabilities, and allow it to address a much wider range of science questions. Similar to the Chandra grating spectrometers the instrument could be folded into the beam only when needed or could be permanently inserted. As transmission gratings are largely transparent at high energies, in this latter case hard X-rays could still be used simultaneously by the prime instrument.
Soft X-ray grating spectroscopy naturally pairs with UV observations since the UV covers tracers of gas in similar temperature ranges. Either a UV spectrograph or an UV imager or both can be added. Those instruments would use separate, but aligned, optics and essentially be bolted onto the outside of the X-ray mission, like the OM on XMM-Newton, so this concept is a proven design. 

\begin{SCfigure}
    \includegraphics[width=0.6\textwidth]{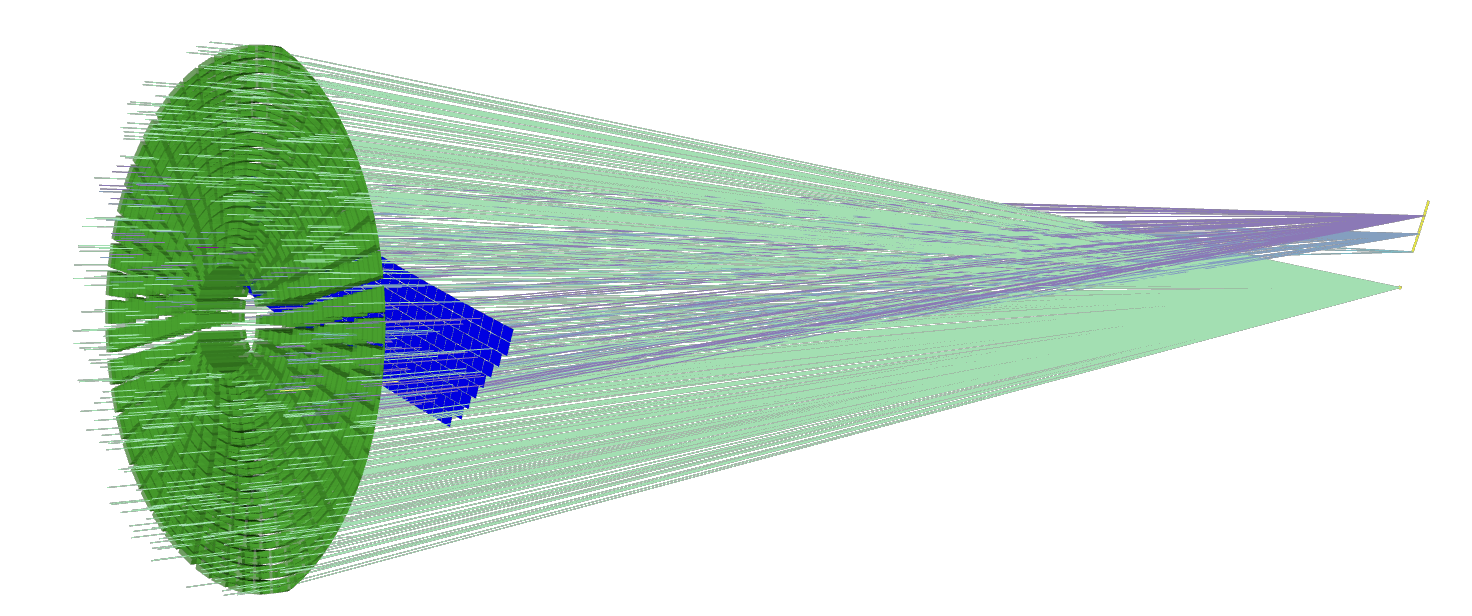}
    \caption{Instrument consisting of a grating petal (blue) that is folded in behind the mirror of the host mission (green). When folded in, it disperses some photons (blue and purple) onto a dedicated strip of detectors.}
    \label{fig:instrument}
\end{SCfigure}

\textbf{X-ray grating spectrometer:} Development of critical-angle transmission gratings (CAT gratings) with much deeper bars than on Chandra has raised grating efficiency significantly. Depending on the form factor of the grating and any coating used, $>$ 20-40\% efficiency can be achieved in the $\sim$ 0.2-2 keV range, with a resolving power of $R > 10000$ at 1 keV \cite{2025JATIS..11a1004H}; in other words, the resolving power of the instrument is limited by the optics of the host mission, not by the gratings themselves. The geometry is a Rowland torus, where the gratings are placed shortly behind the focussing optic and a detector on the opposite side of the torus \cite{1978ApOpt..17.2304B}, see Fig~\ref{fig:instrument}. In blazed CAT gratings, the grating spectra are dispersed from the direct light by $3-5^{\circ}$, corresponding to a distance of 30-80~cm from the prime instrument for focal lengths of 5-10~m. At that location, the instrument needs a strip of detectors (about $10\times300$~mm). Detector requirements are modest (read-out times of order 0.1-1~s, intrinsic energy resolution similar to 30 year-old Chandra CCDs, pixel size $\leq$50~$\mu$m). Either CCD or CMOS devices will do. For simplicity and cost savings, the detector type can be chosen to match the prime instrument if applicable or off-the-shelf devices can be used, keeping risk and cost low.

Details for this instrument obviously depend on the host mission, but design studies and ray-tracing show this conept works for the $1^{\prime\prime}$ optics on Lynx (reaching $R>5000$ \cite{2019JATIS...5b1003G} for a soft X-ray grating spectrograph) or AXIS ($R=4000$ with effective area 1500~cm$^2$ or $R= 6000$ with 500~cm$^2$ \cite{2023JATIS...9b4007G}), as well as with Athena's $5^{\prime\prime}$ optics (reaching $R>3000$ \cite{2025JATIS..11a1005G}). In either case, the spectrometer design will take advantage of sub-aperturing, i.e.,\ covering only a fraction of the mirror area to increase the spectral resolution.

Gratings are manufactured using lithography and etch processes well established in the semiconductor and MEMS industries, and large wafers are then etched and cut into individual gratings. Gratings of size $30\times30\;\mathrm{mm}^2$ have reached TRL 5 or 6 in the development for the Arcus mission concepts, and smaller CAT gratings will be flown in 2028 on REDSoX and in 2030 on GOSoX. We can increase the throughput by reducing the width of the support structure and increasing the grating size. Some development is needed to reach TRL 6 in this scenario, depending on the science goals and the X-ray optic; essentially, one needs to write a different mask for the UV lithography and repeat environmental testing. Further development to coat the grating bars, bend gratings, or chirp the period of gratings bars within a grating would enable even higher efficiency and resolution while reducing cost of manufacturing by reducing the number of grating needed.

\textbf{UV instruments}
An UV imager with sufficient capabilities has flown on GALEX (the instrument itself costs only a fraction of the SMEX budget) or could be a scaled-down version of the UVEX instrument that is already under development. An UV spectrograph with 60~cm diameter aperture has been developed for the Arcus Probe concept based on experience with HST/COS \cite{2023SPIE12678E..0GF,2025JATIS..11a1002F}.

\textbf{Cost} The addition of X-ray grating and UV instruments would add of order \$100-200M to the cost of the host mission.  Reusing existing instrumentation, such as the X-ray focal plane detectors or electronics, could result in substantial cost savings, as can design choices dependent on the host mission configuration.  This is consistent with cost estimates for the Arcus and Lynx proposals, both of which went through detailed budgeting exercises.

\section{Mission Implementation} % < 1 page

%Since the X-ray grating spectrometer is slit-less, targets will be mostly isolated point sources (AGN, X-ray binaries, stars). 

\textbf{The main idea behind the concept presented here is to include high-resolution soft X-ray spectroscopy and UV imaging/spectroscopy into the design of another X-ray mission.} This approach vastly enhances the science return of the host mission at a small fraction of the cost of a stand-alone mission. Both of the most successful X-ray missions ever launched, Chandra and XMM-Newton, included X-ray grating spectrometers as part of a mission that performs imaging most of the time. XMM-Newton also includes a separate small, but successful optical/UV telescope similar to what we propose here. 

A fraction for the observing time of the host mission would be allocated to grating spectroscopy; for Chandra and XMM-Newton this is done in the peer review process, which on average allocates about 10\% of the available time to gratings observations. With a similar fraction of observation time, the science questions posed in Section 2 can be answered in a 5-year mission.

X-ray grating spectrometer and UV instruments are very flexible and can be added to X-ray missions with focussing mirrors independent of the host missions orbit, launch vehicle, etc. The grating specific requirement is a pointing stability $<1.5^{\prime\prime}$/s or so (likely to be achieved by any major X-ray mission today). For the X-ray grating spectrometer data rates (600~kb/s), power ($<200$~W) and instrument mass ($<200$~kg, readouts and grating arrays combined) are each only a small fraction of the resources of a major X-ray mission. (Shown numbers are conservative estimates derived from the Lynx CSR with $R > 5000$ and $A_{eff} > 4000$ cm$^2$.) 

Unless the host mission already uses CCDs or CMOSs (in which case the same type can be used for the grating spectrometer), the X-ray instrument could leverage existing sensors from a range of vendors, including off-the-shelf sensors from commercial vendors with minimal development for packaging and electronics.

UV imaging can be done with a small ($\approx 30$~cm) telescope, with very low weight and power requirements. Unlike the OM on XMM, the data product can be an integrated image to keep data rates low if needed. 
A UV spectrograph would add more weight than a UV imager because it requires a larger telescope, but data rates and power use can be limited by choosing a narrower bandpass to accommodate any limits imposed by the architecture of the host mission.

Most observations of in the science table can be done without timing constraints, but some targets profit from either regular spectroscopic monitoring or longer observations with limited interruptions. The UV instruments will simultaneously acquire images and spectra of the same target. Data processing and analysis is well-developed from Chandra HETG and LETG and XMM-Newton/RGS. Algorithms and even parts of the reduction software can be reused. Several software packages to analyse and fit X-ray spectra are widely used in the community and are suitable to fit the data from the proposed instrument. Some additional laboratory and theoretical work on dust absorption can improve the science return.

\DeclareRobustCommand{\ion}[2]{%
\relax\ifmmode
\ifx\testbx\f@series
{\mathbf{#1\,\mathsc{#2}}}\else
{\mathrm{#1\,\mathsc{#2}}}\fi
\else\textup{#1\,{\mdseries\textsc{#2}}}%
\fi}

\def\farcm{\hbox{$.\mkern-4mu^\prime$}}
\def\farcs{\hbox{$.\!\!^{\prime\prime}$}}
\def\degr{\hbox{$^\circ$}}
\def\arcmin{\hbox{$^\prime$}}
\def\arcsec{\hbox{$^{\prime\prime}$}}

\def\na{New Astronomy}
\def\rmxaa{Revista Mexicana de Astronomía y Astrofísica}
\def\aj{AJ}%

          % Astronomical Journal

\def\araa{ARA\&A}%

          % Annual Review of Astron and Astrophys

\def\apj{ApJ}%

          % Astrophysical Journal

\def\apjl{ApJ}%

          % Astrophysical Journal, Letters

\def\apjs{ApJS}%

          % Astrophysical Journal, Supplement

\def\ao{Appl.~Opt.}%

          % Applied Optics

\def\apss{Ap\&SS}%

          % Astrophysics and Space Science

\def\aap{A\&A}%

          % Astronomy and Astrophysics

\def\aapr{A\&A~Rev.}%

          % Astronomy and Astrophysics Reviews

\def\aaps{A\&AS}%

          % Astronomy and Astrophysics, Supplement

\def\azh{AZh}%

          % Astronomicheskii Zhurnal

\def\baas{BAAS}%

          % Bulletin of the AAS

\def\jrasc{JRASC}%

          % Journal of the RAS of Canada

\def\memras{MmRAS}%

          % Memoirs of the RAS

\def\mnras{MNRAS}%

          % Monthly Notices of the RAS

\def\pra{Phys.~Rev.~A}%

          % Physical Review A: General Physics

\def\prb{Phys.~Rev.~B}%

          % Physical Review B: Solid State

\def\prc{Phys.~Rev.~C}%

          % Physical Review C

\def\prd{Phys.~Rev.~D}%

          % Physical Review D

\def\pre{Phys.~Rev.~E}%

          % Physical Review E

\def\prl{Phys.~Rev.~Lett.}%

          % Physical Review Letters

\def\pasp{PASP}%

          % Publications of the ASP

\def\pasj{PASJ}%

          % Publications of the ASJ

\def\qjras{QJRAS}%

          % Quarterly Journal of the RAS

\def\skytel{S\&T}%

          % Sky and Telescope

\def\solphys{Sol.~Phys.}%

          % Solar Physics

\def\sovast{Soviet~Ast.}%

          % Soviet Astronomy

\def\ssr{Space~Sci.~Rev.}%

          % Space Science Reviews

\def\zap{ZAp}%

          % Zeitschrift fuer Astrophysik

\def\nat{Nature}%

          % Nature

\def\iaucirc{IAU~Circ.}%

          % IAU Cirulars

\def\aplett{Astrophys.~Lett.}%

          % Astrophysics Letters

\def\apspr{Astrophys.~Space~Phys.~Res.}%

          % Astrophysics Space Physics Research

\def\bain{Bull.~Astron.~Inst.~Netherlands}%

          % Bulletin Astronomical Institute of the Netherlands

\def\fcp{Fund.~Cosmic~Phys.}%

          % Fundamental Cosmic Physics

\def\gca{Geochim.~Cosmochim.~Acta}%

          % Geochimica Cosmochimica Acta

\def\grl{Geophys.~Res.~Lett.}%

          % Geophysics Research Letters

\def\jcp{J.~Chem.~Phys.}%

          % Journal of Chemical Physics

\def\jgr{J.~Geophys.~Res.}%

          % Journal of Geophysics Research

\def\jqsrt{J.~Quant.~Spec.~Radiat.~Transf.}%

          % Journal of Quantitiative Spectroscopy and Radiative Trasfer

\def\memsai{Mem.~Soc.~Astron.~Italiana}%

          % Mem. Societa Astronomica Italiana

\def\nphysa{Nucl.~Phys.~A}%

          % Nuclear Physics A

\def\physrep{Phys.~Rep.}%

          % Physics Reports

\def\physscr{Phys.~Scr}%

          % Physica Scripta

\def\planss{Planet.~Space~Sci.}%

          % Planetary Space Science

\def\procspie{Proc.~SPIE}%

          % Proceedings of the SPIE

\def\nar{New Astronomy Reviews}

\let\astap=\aap

\let\apjlett=\apjl

\let\apjsupp=\apjs

\let\applopt=\ao

\uchyph=0

\setlength\bibsep{0pt}
\bibliography{references}

\begin{thebibliography}{23}
\expandafter\ifx\csname natexlab\endcsname\relax\def\natexlab#1{#1}\fi

\bibitem[{{Anderson} \& {Bregman}(2011)}]{2011ApJ...737...22A}
{Anderson} \& {Bregman} 2011, \apj, 737, 22

\bibitem[{{Beuermann} {et~al.}(1978){Beuermann}, {Braeuninger}, \&
  {Truemper}}]{1978ApOpt..17.2304B}
{Beuermann} {et~al.} 1978, \ao, 17, 2304

\bibitem[{{Bregman} {et~al.}(2015){Bregman}, {Alves}, {Miller}, \&
  {Hodges-Kluck}}]{2015JATIS...1d5003B}
{Bregman} {et~al.} 2015, Journal of Astronomical Telescopes, Instruments, and
  Systems, 1, 045003

\bibitem[{{Davies} {et~al.}(2020){Davies}, {Crain}, {Oppenheimer}, \&
  {Schaye}}]{2020MNRAS.491.4462D}
{Davies} {et~al.} 2020, \mnras, 491, 4462

\bibitem[{{Ferrarese} \& {Merritt}(2000)}]{2000ApJ...539L...9F}
{Ferrarese} \& {Merritt} 2000, \apjl, 539, L9

\bibitem[{{Fleming} {et~al.}(2023){Fleming}, {France}, {Patton}, {Hellickson},
  {Nell}, {Smith}, \& {Cheimets}}]{2023SPIE12678E..0GF}
{Fleming} {et~al.} 2023, in Society of Photo-Optical Instrumentation Engineers
  (SPIE) Conference Series, Vol. 12678, UV, X-Ray, and Gamma-Ray Space
  Instrumentation for Astronomy XXIII, ed. O.~H. {Siegmund} \& K.~{Hoadley},
  126780G

\bibitem[{{France} {et~al.}(2025){France}, {Fleming}, {Brenneman}, {Smith},
  {Bregman}, {Brickhouse}, {G{\"u}nther}, {Tripp}, {Bhattacharyya},
  {Hellickson}, {Nell}, {Patton}, {Poppenhaeger}, \&
  {Temi}}]{2025JATIS..11a1002F}
{France} {et~al.} 2025, Journal of Astronomical Telescopes, Instruments, and
  Systems, 11, 011002

\bibitem[{{Fukugita} \& {Peebles}(2004)}]{2004ApJ...616..643F}
{Fukugita} \& {Peebles} 2004, \apj, 616, 643

\bibitem[{{G{\"u}nther} {et~al.}(2025){G{\"u}nther}, {Cheimets}, {DeRoo}, \&
  {Heilmann}}]{2025JATIS..11a1005G}
{G{\"u}nther} {et~al.} 2025, Journal of Astronomical Telescopes, Instruments,
  and Systems, 11, 011005

\bibitem[{{G{\"u}nther} \& {Heilmann}(2019)}]{2019JATIS...5b1003G}
{G{\"u}nther} \& {Heilmann} 2019, Journal of Astronomical Telescopes,
  Instruments, and Systems, 5, 021003

\bibitem[{{G{\"u}nther} {et~al.}(2023){G{\"u}nther}, {Principe}, {Heilmann},
  {Cheimets}, {Hertz}, \& {Smith}}]{2023JATIS...9b4007G}
{G{\"u}nther} {et~al.} 2023, Journal of Astronomical Telescopes, Instruments,
  and Systems, 9, 024007

\bibitem[{{Heilmann} {et~al.}(2025){Heilmann}, {Bruccoleri}, {Gregory},
  {Gullikson}, {G{\"u}nther}, {Hertz}, {Lambert}, {Young}, \&
  {Schattenburg}}]{2025JATIS..11a1004H}
{Heilmann} {et~al.} 2025, Journal of Astronomical Telescopes, Instruments, and
  Systems, 11, 011004

\bibitem[{{Huber} \& {Bregman}(2022)}]{2022AJ....163..264H}
{Huber} \& {Bregman} 2022, \aj, 163, 264

\bibitem[{{King} \& {Pounds}(2015)}]{2015ARA&A..53..115K}
{King} \& {Pounds} 2015, \araa, 53, 115

\bibitem[{{Li} \& {Bregman}(2017)}]{2017ApJ...849..105L}
{Li} \& {Bregman} 2017, \apj, 849, 105

\bibitem[{{Magorrian} {et~al.}(1998){Magorrian}, {Tremaine}, {Richstone},
  {Bender}, {Bower}, {Dressler}, {Faber}, {Gebhardt}, {Green}, {Grillmair},
  {Kormendy}, \& {Lauer}}]{1998AJ....115.2285M}
{Magorrian} {et~al.} 1998, \aj, 115, 2285

\bibitem[{{Onaka} {et~al.}(2025){Onaka}, {Sakon}, {Shimonishi}, \&
  {Honda}}]{2025prim.book..446O}
{Onaka} {et~al.} 2025, in PRIMA General Observer Science Book Volume 2, ed.
  A.~{Moullet}, D.~{Burgarella}, T.~{Kataria}, H.~{Beuther}, C.~{Battersby},
  M.~{Cheng}, T.~{Essinger-Hileman}, H.~{Inami}, E.~{Mills}, T.~{Nagao}, \&
  S.~{Unwin}, Vol.~2, 446--451

\bibitem[{{Planck Collaboration} {et~al.}(2013){Planck Collaboration}, {Ade},
  {Aghanim}, {Arnaud}, {Ashdown}, {Atrio-Barandela}, {Aumont}, {Baccigalupi},
  {Balbi}, {Banday}, {Barreiro}, {Barrena}, {Bartlett}, {Battaner}, {Benabed},
  {Bernard}, {Bersanelli}, {Bikmaev}, {Bock}, {B{\"o}hringer}, {Bonaldi},
  {Bond}, {Borrill}, {Bouchet}, {Bourdin}, {Burenin}, {Burigana}, {Butler},
  {Cabella}, {Chamballu}, {Chary}, {Chiang}, {Chon}, {Christensen}, {Clements},
  {Colafrancesco}, {Colombi}, {Colombo}, {Comis}, {Coulais}, {Crill},
  {Cuttaia}, {Da Silva}, {Dahle}, {Davis}, {de Bernardis}, {de Gasperis}, {de
  Rosa}, {de Zotti}, {Delabrouille}, {D{\'e}mocl{\`e}s}, {Diego}, {Dole},
  {Donzelli}, {Dor{\'e}}, {Douspis}, {Dupac}, {Efstathiou}, {En{\ss}lin},
  {Finelli}, {Flores-Cacho}, {Forni}, {Frailis}, {Franceschi}, {Frommert},
  {Galeotta}, {Ganga}, {G{\'e}nova-Santos}, {Giard}, {Giraud-H{\'e}raud},
  {Gonz{\'a}lez-Nuevo}, {G{\'o}rski}, {Gregorio}, {Gruppuso}, {Hansen},
  {Harrison}, {Hern{\'a}ndez-Monteagudo}, {Herranz}, {Hildebrandt}, {Hivon},
  {Hobson}, {Holmes}, {Hornstrup}, {Hovest}, {Huffenberger}, {Hurier}, {Jaffe},
  {Jaffe}, {Jones}, {Juvela}, {Keih{\"a}nen}, {Keskitalo}, {Khamitov},
  {Kisner}, {Kneissl}, {Knoche}, {Kunz}, {Kurki-Suonio}, {L{\"a}hteenm{\"a}ki},
  {Lamarre}, {Lasenby}, {Lawrence}, {Le Jeune}, {Leonardi}, {Lilje},
  {Linden-V{\o}rnle}, {L{\'o}pez-Caniego}, {Lubin}, {Luzzi},
  {Mac{\'\i}as-P{\'e}rez}, {MacTavish}, {Maffei}, {Maino}, {Mandolesi},
  {Maris}, {Marleau}, {Marshall}, {Mart{\'\i}nez-Gonz{\'a}lez}, {Masi},
  {Massardi}, {Matarrese}, {Mazzotta}, {Mei}, {Melchiorri}, {Melin}, {Mendes},
  {Mennella}, {Mitra}, {Miville-Desch{\^e}nes}, {Moneti}, {Montier},
  {Morgante}, {Mortlock}, {Munshi}, {Murphy}, {Naselsky}, {Nati}, {Natoli},
  {N{\o}rgaard-Nielsen}, {Noviello}, {Novikov}, {Novikov}, {Osborne},
  {Oxborrow}, {Pajot}, {Paoletti}, {Perotto}, {Perrotta}, {Piacentini}, {Piat},
  {Pierpaoli}, {Piffaretti}, {Plaszczynski}, {Pointecouteau}, {Polenta},
  {Popa}, {Poutanen}, {Pratt}, {Prunet}, {Puget}, {Rachen}, {Rebolo},
  {Reinecke}, {Remazeilles}, {Renault}, {Ricciardi}, {Ristorcelli}, {Rocha},
  {Roman}, {Rosset}, {Rossetti}, {Rubi{\~n}o-Mart{\'\i}n}, {Rusholme},
  {Sandri}, {Savini}, {Scott}, {Spencer}, {Starck}, {Stolyarov}, {Sudiwala},
  {Sunyaev}, {Sutton}, {Suur-Uski}, {Sygnet}, {Tauber}, {Terenzi},
  {Toffolatti}, {Tomasi}, {Tristram}, {Valenziano}, {Van Tent}, {Vielva},
  {Villa}, {Vittorio}, {Wade}, {Wandelt}, {Wang}, {Welikala}, {Weller}, \&
  {White}}]{2013A&A...557A..52P}
{Planck Collaboration} {et~al.} 2013, \aap, 557, A52

\bibitem[{{Psaradaki} {et~al.}(2020){Psaradaki}, {Costantini}, {Mehdipour},
  {Rogantini}, {de Vries}, {de Groot}, {Mutschke}, {Trasobares}, {Waters}, \&
  {Zeegers}}]{2020A&A...642A.208P}
{Psaradaki} {et~al.} 2020, \aap, 642, A208

\bibitem[{{Psaradaki} {et~al.}(2023){Psaradaki}, {Costantini}, {Rogantini},
  {Mehdipour}, {Corrales}, {Zeegers}, {de Groot}, {den Herder}, {Mutschke},
  {Trasobares}, {de Vries}, \& {Waters}}]{2023A&A...670A..30P}
{Psaradaki} {et~al.} 2023, \aap, 670, A30

\bibitem[{{Rogantini} {et~al.}(2026){Rogantini}, {Canizares}, {Costantini},
  {Gu}, {Mehdipour}, {Psaradaki}, {Schulz}, \& {Zeegers}}]{2026ApJ..1002...86R}
{Rogantini} {et~al.} 2026, \apj, 1002, 86

\bibitem[{{Rogantini} {et~al.}(2018){Rogantini}, {Costantini}, {Zeegers}, {de
  Vries}, {Bras}, {de Groot}, {Mutschke}, \& {Waters}}]{2018A&A...609A..22R}
{Rogantini} {et~al.} 2018, \aap, 609, A22

\bibitem[{{Tombesi} {et~al.}(2010){Tombesi}, {Cappi}, {Reeves}, {Palumbo},
  {Yaqoob}, {Braito}, \& {Dadina}}]{2010A&A...521A..57T}
{Tombesi} {et~al.} 2010, \aap, 521, A57

\end{thebibliography}
\end{document}